\documentclass[12pt,preprint]{aastex}

\shorttitle{Short-lived radionuclides from a faint supernova with mixing-fallback}
\shortauthors{Takigawa et al.}

\begin{document}

\title{Injection of Short-Lived Radionuclides into the Early Solar System from a Faint Supernova with Mixing-Fallback}

\author{A. Takigawa\altaffilmark{1}, J. Miki\altaffilmark{1}, S. Tachibana\altaffilmark{1},  G. R. Huss\altaffilmark{2},  N. Tominaga\altaffilmark{3}, H. Umeda\altaffilmark{3}, and K. Nomoto\altaffilmark{3,4,5}}

\altaffiltext{1}{Department of Earth and Planetary Science, University of Tokyo, 7-3-1 Hongo, Tokyo 113-0033, Japan; takigawa@eps.s.u-tokyo.ac.jp, miki@eps.s.u-tokyo.ac.jp, tachi@eps.s.u-tokyo.ac.jp.}
\altaffiltext{2}{Hawaii Institute of Geophysics and Planetology, University of Hawaii at Manoa, 1680 East-West Road, Honolulu, HI 96822; ghuss@higp.hawaii.edu.}
\altaffiltext{3}{Department of Astronomy, University of Tokyo, 7-3-1 Hongo, Tokyo 113-0033, Japan; tominaga@astron.s.u-tokyo.ac.jp, umeda@astron.s.u-tokyo.ac.jp, nomoto@astron.s.u-tokyo.ac.jp.}
\altaffiltext{4}{Research Center for the Early Universe, University of Tokyo, 7-3-1 Hongo, Tokyo 113-0033, Japan}
\altaffiltext{5}{Institute for the Physics and Mathematics of the Universe, University of Tokyo, 5-1-5 Kashiwa-no-ha, Kashiwa, Chiba 277-8581, Japan}

\begin{abstract}
Several short-lived radionuclides (SLRs) were present in the early solar system, some of which should have formed just prior to or soon after the solar system formation.  Stellar nucleosynthesis has been proposed as the mechanism for production of SLRs in the solar system, but no appropriate stellar source has been found to explain the abundances of all solar system SLRs.

In this study, we propose a faint supernova with mixing and fallback as a stellar source of SLRs with mean lives of $<$5 Myr ($^{26}$Al, $^{41}$Ca, $^{53}$Mn, and $^{60}$Fe) in the solar system.  In such a supernova, the inner region of the exploding star experiences mixing, a small fraction of mixed materials is ejected, and the rest undergoes fallback onto the core.  The modeled SLR abundances agree well with their solar system abundances if mixing-fallback occurs within the C/O-burning layer.  In some cases, the initial solar system abundances of the SLRs can be reproduced within a factor of 2.  The dilution factor of supernova ejecta to the solar system materials is $\sim$10$^{-4}$ and the time interval between the supernova explosion and the formation of oldest solid materials in the solar system is $\sim$1 Myr.  If the dilution occurred due to spherically symmetric expansion, a faint supernova should have occurred nearby the solar system forming region in a star cluster.
\end{abstract}

\section{Introduction}

The former presence of short-lived radionuclides (SLRs) in the early solar system ($^{10}$Be, $^{26}$Al, $^{36}$Cl, $^{41}$Ca, $^{53}$Mn, $^{60}$Fe, $^{107}$Pd, $^{129}$I, and $^{182}$Hf) has been inferred from excesses in the abundances of their daughter nuclides in meteorites, linearly correlated with the abundance of a parent element \citep[e.g.,][]{McD03,Ki05}.  These now-extinct SLRs may provide high-resolution ($<$0.1 million years (Myr)) chronometers for events that occurred during the first several million years of solar system evolution and may also be potential heat sources for asteroidal metamorphism and/or differentiation.  

The SLRs with relatively long mean-lives such as $^{107}$Pd, $^{129}$I, $^{182}$Hf, and perhaps $^{53}$Mn may have been products of steady-state nucleosynthesis in the galaxy \citep{Ja05}, while those with mean-lives ($\tau$) of $<$5 Myr, $^{10}$Be ($\tau$=2.2 Myr), $^{26}$Al ($\tau$=1.03 Myr), $^{36}$Cl ($\tau$=0.43 Myr), $^{41}$Ca ($\tau$=0.15 Myr), $^{60}$Fe ($\tau$=2.2 Myr), and possibly $^{53}$Mn ($\tau$=5.3 Myr), should have been produced either by energetic-particle irradiation in the early solar system or by stellar nucleosynthesis just prior to or shortly after the birth of the solar system.  Beryllium-10, which was found in CAIs (calcium-aluminum-rich inclusions) (McKeegan et al., 2000), is not produced by stellar nucleosynthesis, but can be efficiently formed by energetic particle irradiation.  On the other hand, $^{60}$Fe, the abundance of which in the early solar system also appears to require a late addition \citep{TH03,Mo05,Ta06}, can be efficiently produced only in stars.  Thus, the presence of $^{10}$Be and $^{60}$Fe in the early solar system suggests that both energetic particle irradiation and stellar nucleosynthesis make contributions to the inventories of solar system SLRs.  However, it is not clear yet which process contributed more significantly to the inventory of the SLRs that could be synthesized by both processes, such as $^{26}$Al, $^{36}$Cl, $^{41}$Ca, and $^{53}$Mn.

There have been several attempts to find a plausible stellar source(s) for the abundances of SLRs in the early solar system.  A low-mass (1.5 M$_{\sun}$) thermally-pulsing asymptotic-giant-branch (TP-AGB) star cannot produce enough $^{60}$Fe to match the initial abundance of $^{60}$Fe in the solar system \citep[e.g.,][]{Bu03,Wa06,SS06}.  Models for intermediate-mass AGB stars (5 M$_{\sun}$ with the solar metallicity or 3 M$_{\sun}$ with 1/3 $\times$ the solar metallicity) could explain the inferred solar system abundance of $^{60}$Fe as well as the abundances of $^{26}$Al, $^{41}$Ca and $^{107}$Pd \citep{Wa06}.  However, the probability of encounters between molecular clouds and AGB stars seems to be extremely low \citep{KM94}, implying that an AGB star was an unlikely source of SLRs in the solar system.  Mass-loss winds from Wolf-Rayet stars may have been a source of the solar system $^{26}$Al, $^{36}$Cl, $^{41}$Ca, and $^{107}$Pd if the nuclides were incorporated into the solar system within a time-interval of $\sim$10$^{5}$ year after production \citep{Ar06}.  However, a WR wind alone would not contain enough $^{60}$Fe and $^{53}$Mn to explain their solar system initial abundances \citep{Ar06}.  

Type II core-collapse supernovae have also been considered as plausible sources for SLRs.  However, most supernova models imply that if a supernova provided $^{26}$Al and $^{41}$Ca into the solar system, it would also supply 10-100 times more $^{53}$Mn than its estimated initial abundance in the solar system \citep[e.g.,][]{GV00,Ou05,SS06}.  This discrepancy could be explained by fallback of the innermost layers containing most of the $^{53}$Mn onto a collapsing stellar core \citep{MC00,Me05}.  In such a case, $^{53}$Mn could primarily be derived from a different source such as the interstellar medium.  \cite{Wa06} proposed that $^{53}$Mn and possibly $^{60}$Fe were injected into the solar system as supernova ejecta with a time-interval of $\sim$10$^{7}$ years after production, long enough for $^{41}$Ca and $^{26}$Al to decay completely, and that $^{26}$Al and $^{41}$Ca in the solar system may have been produced either by energetic particle irradiation or by an AGB star.

Another problem with supernovae as sources of SLRs is overproduction of $^{60}$Fe if all the $^{26}$Al in the solar system was derived from supernovae.  Although the yield of $^{60}$Fe depends the mass loss and initial mass \citep[e.g.,][]{LC06}, the expected amount of $^{60}$Fe injected from a supernova would be, in general, a few times higher than its highest estimate for the early solar system, as we will show later.  This problem may still remain even in models for fallback supernovae.

In this study, we propose a supernova with mixing and fallback, with a kinetic energy of explosion slightly less than that for a typical supernova ($\sim$10$^{51}$ erg) as a potential source of $^{26}$Al, $^{41}$Ca, $^{53}$Mn, and $^{60}$Fe in the early solar system.  Faint supernovae such as SN1997D and SN1999br have such kinetic energies \citep[e.g.,][]{No06}.  In models for supernovae with mixing-fallback, the inner region of the exploding star experiences mixing, some fraction of mixed materials is ejected, and the rest undergoes fallback onto the core \citep[e.g.,][]{UN02,UN05,No06,To07}.  Nucleosynthesis in a faint supernova with mixing-fallback successfully reproduces the elemental abundance patterns of hyper metal-poor stars \citep{UN03,Iw05,No06}. 
\section{Injection of supernova ejecta into the solar system materials}

The abundance of a SLR injected into preexisting solar system materials can be expressed as follows, assuming that injected materials are well-mixed with preexisting materials \citep[e.g.,][]{Wa06,SS06}.:

\begin{equation}
\frac{N^{\rm SLR}}{N^{\rm SI}}=\frac{N^{\rm SLR}_{\rm ejecta} f_{\rm 0} e^{-\Delta/\tau}}{N^{\rm SI}_{\rm 0}+N^{\rm SI}_{\rm ejecta} f_{\rm 0}}
\end{equation}
where \textit{N}$^{\rm SLR}$ and \textit{N}$^{\rm SI}$ are the numbers of SLR and a stable isotope (SI) for the initial solar system, respectively.  Time-zero for the solar system is, in practice, considered to be the time of formation of CAIs, the oldest solid materials formed in the solar system.  \textit{N}$^{\rm SLR}_{\rm ejecta}$ and \textit{N}$^{\rm SI}_{\rm ejecta}$ are the numbers of SLR and SI at the time of their production, respectively. \textit{N}$^{\rm SI}_{\rm 0}$ is the number of SI in the preexisting solar system materials.  We obtained \textit{N}$^{\rm SLR}_{\rm ejecta}$ and \textit{N}$^{\rm SI}_{\rm ejecta}$ for supernovae with mixing-fallback based on nucleosynthesis models by Nomoto et al. (2006).  Their values depend on variable parameters for mixing-fallback such as the scale of the mixing region and an ejection efficiency of mixed materials.  Details of determination of \textit{N}$^{\rm SLR}_{\rm ejecta}$ and \textit{N}$^{\rm SI}_{\rm ejecta}$ ejecta in the context of a mixing-fallback model are shown below. \textit{N}$^{\rm SLR}_{\rm ejecta}$ and \textit{N}$^{\rm SI}_{\rm ejecta}$ were also obtained for non-fallback and fallback supernovae using nucleosynthesis models by \cite{WW95}, \cite{Ra02}, \cite{CL04}, and \cite{No06} for comparison.  \textit{N}$^{\rm SI}_{\rm 0}$ was taken from \cite{AG89}.  The variables \textit{f}$_{\rm 0}$ and $\Delta$ are free parameters in the injection model representing a dilution factor for supernova ejecta mixed with the preexisting materials and the time interval between the production of SLR and the CAI formation, respectively.
\section{Nuclides ejected from supernovae with mixing-fallback}

In the standard model for a supernova almost all the materials are ejected, but materials in the very inner-most region, which is typically deeper than the incomplete Si-burning layer, fall back onto the star.  In the model for a supernova with fallback \citep{MC00,Me05}, such a mass-cut boundary was moved out to the incomplete Si-burning layer or even out to the C/O burning layer, which resulted in suppressed ejection of $^{53}$Mn.

In the model for a supernova with mixing-fallback, we assume two mass-cut boundaries at different depths of a pre-supernova star.  The deeper boundary is defined as the initial mass-cut boundary (\textit{M}$_{\rm cut}$) corresponding to the mass-cut boundary in previous models. The shallower boundary is defined as the outer boundary of the mixing region (\textit{M}$_{\rm mix}$), which is considered to be the volume between \textit{M}$_{\rm cut}$ and \textit{M}$_{\rm mix}$.  Most of materials in the region between \textit{M}$_{\rm cut}$ and \textit{M}$_{\rm mix}$ fallback onto the core by gravity after complete mixing by Rayleigh-Taylor instabilities, but a small fraction (\textit{q}) of homogeneously-mixed materials is ejected from the mixing region.  Thus, in the present model, supernova ejecta consist of all the nuclides from the region outside of \textit{M}$_{\rm mix}$ and a fraction \textit{q} of nuclides within the mixing region between \textit{M}$_{\rm cut}$ and \textit{M}$_{\rm mix}$.

Two mass-cut boundaries (\textit{M}$_{\rm cut}$ and \textit{M}$_{\rm mix}$) and the fraction of materials ejected from the mixing region (\textit{q}) are essential parameters to determine \textit{N}$^{\rm SLR}_{\rm ejecta}$ and \textit{N}$^{\rm SI}_{\rm ejecta}$ \citep[e.g.,][]{Iw05,No06}.  The kinetic energies of explosion of faint supernovae (SN1997D and SN1999br) are estimated to be $\sim$(0.4-0.6)$\times$10$^{51}$ erg, which is $\sim$1/3 to 1/2 times smaller than the typical kinetic energy of supernova explosion.  However, an explosion with slightly less kinetic energy does not significantly affect explosive nucleosyntheses \citep[e.g.,][]{WW95}, and thus we evaluated \textit{N}$^{\rm SLR}_{\rm ejecta}$ and \textit{N}$^{\rm SI}_{\rm ejecta}$ with different sets of (\textit{M}$_{\rm cut}$, \textit{M}$_{\rm mix}$ and \textit{q} based on a nucleosynthesis model for a solar-metallicity massive star with the kinetic energy of explosion of 10$^{51}$ erg (Fig. 1; \cite{No06}).  

Regarding \textit{M}$_{\rm cut}$, the initial mass-cut within the incomplete Si-burning layer suppresses the ejection of $^{53}$Mn as discussed in fallback supernova models \citep{MC00,Me05}.  Although a proper choice of \textit{M}$_{\rm cut}$ in the narrow incomplete Si-burning layer may explain the amounts of SLRs in the solar system including $^{53}$Mn, very precise tuning of the model is required.  In the present study, we varied \textit{M}$_{\rm cut}$ from the bottom of the incomplete Si-burning layer to the deeper region as in standard models to check its effect in the context of mixing-fallback.  We found that the choice of \textit{M}$_{\rm cut}$ affected \textit{N}$^{\rm 56Fe}_{\rm ejecta}$, but did not change significantly \textit{N}$^{\rm SLR}_{\rm ejecta}$ as long as \textit{M}$_{\rm cut}$ was set deeper than the bottom of the incomplete Si-burning region, where no effective nucleosynthesis of SLRs discussed in this study takes place.  As we will show below, \textit{f}$_{\rm 0}$ is estimated to be much smaller than 1, and the change of \textit{N}$^{\rm 56Fe}_{\rm ejecta}$ due to change of the initial mass cut has almost no effects on the following discussion.  We will show results for \textit{M}$_{\rm cut}$ placed at the boundary of the iron core, where peak temperatures are (10-8.5)$\times$10$^{9}$K.

The position of the outer boundary of the mixing region, \textit{M}$_{\rm mix}$, was varied from within the incomplete Si-burning layer to the bottom of H-burning shell, and the ejection fraction, \textit{q}, was given values between 10$^{-4}$-10$^{-2}$.  Note that \textit{M}$_{\rm mix}$ and \textit{q} were required to be within the C/O-burning region or the He-burning region and $\sim$10$^{-4}$, respectively, to explain the elemental abundances of hyper metal-poor stars (HE0107-5240 and HE1327-2326) by a faint supernova \citep{Iw05}.  The upper limit of \textit{q} was determined so that the yield of $^{56}$Ni did not to exceed those for faint supernovae ($\sim$10$^{-3}$ M$_{\sun}$) \citep[e.g.,][]{No06}.   

\section{Modeling the solar system SLRs with a supernova with mixing-fallback }

We determined \textit{f}$_{\rm 0}$ and $\Delta$ to minimize deviations of modeled abundances of $^{26}$Al, $^{41}$Ca, $^{53}$Mn, and $^{60}$Fe (see Eq.(1)) from their estimated initial abundances in the early solar system; $^{26}$Al/$^{27}$Al = (5.0$\pm$0.5)$\times$10$^{-5}$, $^{41}$Ca/$^{40}$Ca = (1.4$\pm$0.14)$\times$10$^{-8}$, $^{53}$Mn/$^{55}$Mn = (9$\pm$4.5)$\times$10$^{-6}$, and $^{60}$Fe/$^{56}$Fe = (7.5$\pm$2.5)$\times$10$^{-7}$ \citep[e.g.,][]{Ki05}.  Because the initial abundances of $^{53}$Mn and $^{60}$Fe have not been well determined yet, we chose plausible abundances for them and gave larger uncertainties than for $^{26}$Al and $^{41}$Ca based on ranges of reported initial abundances.  Note that the choice of plausible initial abundances for $^{53}$Mn and $^{60}$Fe did not have a significant effect on the following discussion.  The uncertainties for the initial abundances were used as weights for minimization of \textit{f}$_{\rm 0}$ and $\Delta$.  Minimization was done in wide varieties of \textit{f}$_{\rm 0}$ and $\Delta$, of which typical steps were 5 \% and 100 years for \textit{f}$_{\rm 0}$ and $\Delta$, respectively, and we confirmed that the grand minimum was obtained. 
\section{Short-lived radionuclides from supernovae with mixing-fallback}

Calculated initial abundances of $^{26}$Al, $^{41}$Ca, $^{53}$Mn, and $^{60}$Fe normalized to meteoritic initial abundances for non-fallback supernovae are shown in Figure 2a-d as a function of the mass of the exploding star.  The dilution factor, \textit{f}$_{\rm 0}$, is estimated to be in the range of (0.5-6)$\times$10$^{-4}$, and typical time-interval, $\Delta$, is $\sim$ 0.7 Myr.  As seen in previous studies, $^{53}$Mn and $^{60}$Fe are, in general, overproduced compared to other SLRs, respectively, irrespective of nucleosynthesis models.  Figure 2e shows the case for fallback models based on the nucleosynthesis model of \cite{WW95} with no mixing, where the amounts of $^{53}$Mn are much less than that expected in the early solar system (after \cite{MC00}).  As mentioned above, if the mass cut occurs within a narrow incomplete Si-burning layer, it might be possible to explain all the SLRs discussed here with fallback supernovae.  However, it requires fine tuning of the mass cut region within the layer. Thus the general consequences of fallback supernova models seems to be very little ejection of $^{53}$Mn.

The abundances of SLRs estimated for faint supernova of different masses with mixing-fallback are shown in Figure 3 as a function of \textit{M}$_{\rm mix}$ and \textit{q}.  It is clearly seen that injection of $^{53}$Mn and $^{60}$Fe is suppressed and the abundances of SLRs agree with their solar system abundances as long as the outer boundary of mixing region is located in the C/O-burning layer, where peak temperatures of the shock wave range from $\sim$4$\times$10$^{8}$ to $\sim$4$\times$10$^{9}$ K.  On the other hand, the modeled abundances of SLRs cannot explain the solar system values if \textit{M}$_{\rm mix}$ is located in the Si-burning or He-burning layers.  \cite{Iw05} proposed that the outer boundary of the mixing layer could be in the upper C/O-burning region or at the bottom of the He-burning layer to explain the elemental abundances of hyper metal-poor stars, HE0107-5240 and HE1327-2326, respectively.  The former case is consistent with \textit{M}$_{\rm mix}$ required to explain the solar system SLRs.  It should be noted here again that the results shown here are not affected by the choice of the initial mass cut(\textit{M}$_{\rm cut}$) as long as it occurs in the region deeper than the incomplete Si-burning layer.

The obtained \textit{f}$_{\rm 0}$ and $\Delta$ increase from 7$\times$10$^{-5}$ to 2$\times$10$^{-3}$ and from 0.8 to 1.1 Myr, respectively, with increasing \textit{M}$_{\rm mix}$ in the C/O-burning layer.  Both parameters affect the abundances of $^{26}$Al and $^{41}$Ca, while those of $^{53}$Mn and $^{60}$Fe are affected mainly by \textit{f}$_{\rm 0}$ when $\Delta$$\sim$1 Myr.  Thus, \textit{f}$_{\rm 0}$ increases with \textit{M}$_{\rm mix}$ because less SLRs are ejected for larger \textit{M}$_{\rm mix}$.  However, the increase of \textit{f}$_{\rm 0}$ to increase the amounts of $^{53}$Mn and $^{60}$Fe may result in the over-injection of $^{26}$Al and $^{41}$Ca, which are adjusted with increasing $\Delta$ in the present model.

In addition of the calculations shown in Fig.3, we calculated the abundances of SLRs with the ejection fraction \textit{q}=10$^{-4}$, which has been proposed for hyper-metal-poor stars \citep{Iw05}, but the abundance of $^{53}$Mn was about ten times lower than the estimate for the initial solar system, as in the simple fallback models.

The best estimates for the faint supernovae with mixing-fallback are shown in Figure 4, where the calculated abundances of SLRs reproduce their solar system abundances within a factor of 2.  It should be emphasized here that a narrow range of mixing-fallback parameters (\textit{M}$_{\rm mix}$ and \textit{q}) is not required to match the solar system abundances of SLRs.  The overall agreement of calculated abundances with meteoritic values can be obtained for wide ranges of \textit{M}$_{\rm mix}$ and \textit{q} (Fig. 3), a major advantage of this model compared to the simple fallback supernova models.
\section{A faint supernova with mixing-fallback as a source of solar system SLRs}

Previous supernova models, which attempted to explain the initial abundances of solar system SLRs, had the problems of over-injection of $^{53}$Mn and $^{60}$Fe compared to $^{26}$Al and $^{41}$Ca.  Such problems can be solved in the faint supernova model with mixing-fallback with the following reasons.

Manganese-53 is synthesized mainly in the incomplete Si-burning layer, while $^{26}$Al, $^{41}$Ca, and $^{60}$Fe are produced more abundantly in outer regions (Fig. 1).  Although a small fraction of the $^{53}$Mn is ejected, most falls back only to the collapsing core when the outer region of the mixing layer is located in the C/O-burning layer.  On the other hand, some fractions of other SLRs, produced outside of the mixing layer, are ejected without experiencing the mixing-fallback.  Together, these features suppress the amount of $^{53}$Mn injected into the solar system materials compared to other SLRs.  

Aluminum-26 is the only nuclide, among SLRs considered in this study, that forms abundantly in the He and H layers and the explosive O-burning layer of a massive star 
with the solar metallicity.  If the oute	r boundary of the mixing region is located in the C/O layer, more than 80 \% of $^{26}$Al should have its origin in the He and H layers.  This is because $^{26}$Al in the C/O layer was destroyed by the He burning \citep[e.g.,][]{LC06} and most of $^{26}$Al in the explosive O-buring layer falls back onto the central remnant. The amount of $^{26}$Al in the supernova ejecta is thus less affected by the mixing-fallback process, which lowers the $^{60}$Fe/$^{26}$Al ratio.  Aluminum-26 may be ejected in a mass-loss wind before the supernova explosion, which may explain the possible time interval between $^{60}$Fe- and $^{26}$Al-injection \citep{Bi07}.

We conclude that $^{26}$Al, $^{41}$Ca, $^{53}$Mn, and $^{60}$Fe in the early solar system could have been brought from a single massive star that experienced a less-energetic supernova explosion.  The estimated dilution factor, \textit{f}$_{\rm 0}$, may provide a geometric constraint on the supernova and the solar system materials as discussed by \cite{SS06}.  Assuming the spherical symmetric ejection from the supernova, \textit{f}$_{\rm 0}$ can be basically expressed by an injection efficiency of ejecta into the solar system materials ($\alpha$) and a solid angle ($\Omega$) of the solar system materials (a proto-solar molecular cloud core or a proto solar system disk) on a sphere of ejecta with a radius corresponding to the distance (\textit{D}) between the supernova and the solar system materials.  If ejecta are injected into the molecular cloud core (0.1 pc in radius, 1 M$_{\sun}$ in mass, and $\alpha$ of 0.1 \citep{VB02}), \textit{D} is estimated to be $\sim$2-5 pc.  In the case of injection into the proto solar system disk as discussed in \cite{Ou05} (100 AU in radius, 0.01 M$_{\sun}$ in mass, and $\alpha$ of 1), \textit{D} of $\sim$0.3-0.8 pc was obtained. 

It is difficult to evaluate which case is more plausible for the solar system forming environment at present.  However, in either case, the supernova explosion should have occurred nearby the solar system materials, which supports the idea of the birth of the solar system in a star-cluster containing massive stars.  The lifetime of the star cluster is several to 10 Myr \citep[e.g.,][]{AL01}, within which only massive stars ($>$20-25 M$_{\sun}$) explode.  The proportion of massive stars in a cluster is low \citep[e.g.,][]{Kr01}, and thus it may be highly unlikely that multiple supernovae brought SLRs into the solar system materials within a lifetime of the star cluster.  In the model for a faint supernova with mixing-fallback, solar system SLRs with mean lives of $<$5 Myr can be from a single supernova.

\acknowledgments

We thank an anonymous reviewer for helpful comments. This work is partly supported by Grant-in-Aid for Scientific Research (18654037), Fujiwara Natural History Foundation (S.T.) and by NASA grants NAG5-11543, NNG05GG48G, and NNX08AG58B (G. R. H.).


\clearpage


\begin{figure}
\epsscale{.80}
\plotone{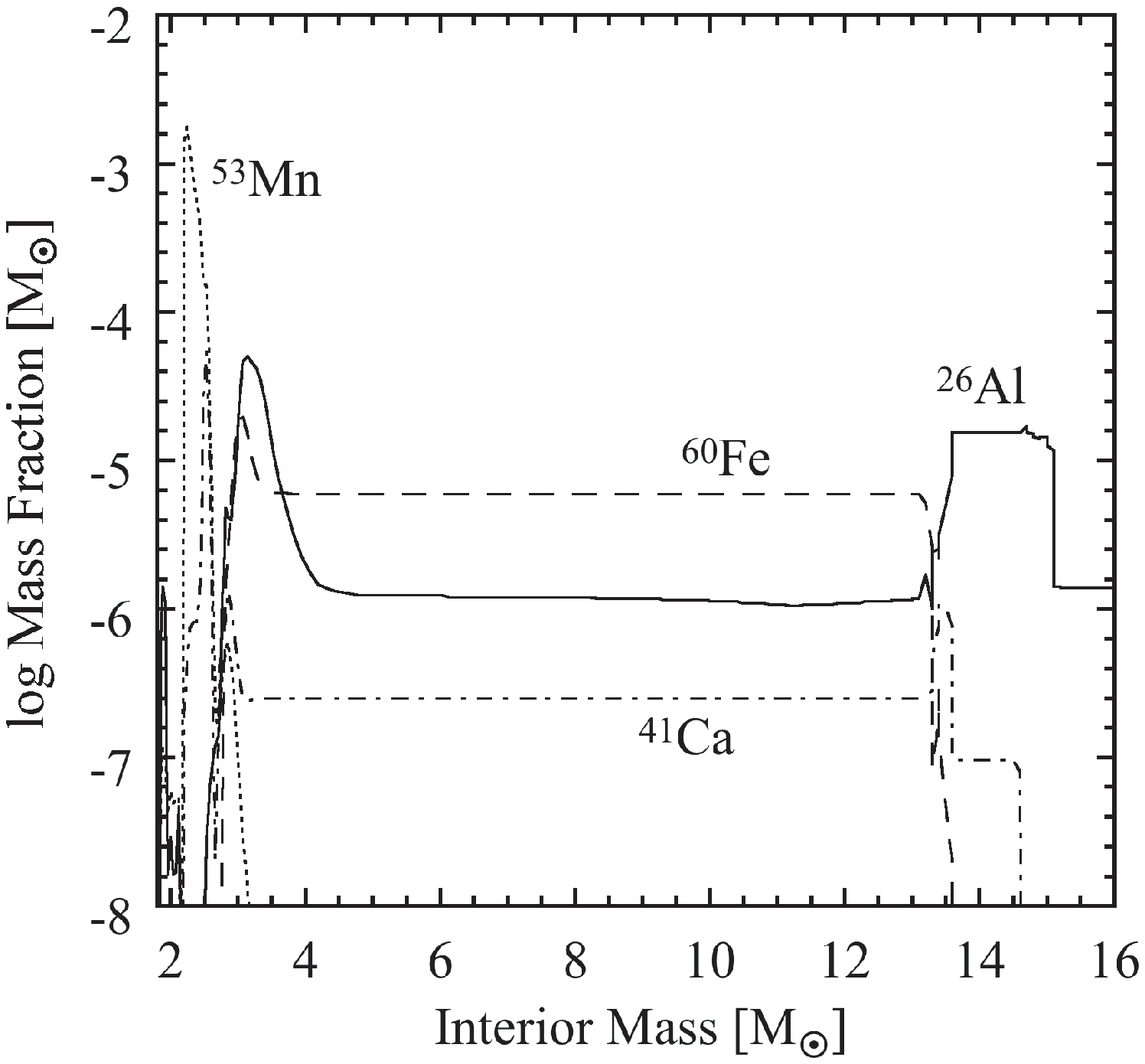}
\caption{Mass fractions of SLRs as a function of Lagrangian mass coordinate after the explosion of a solar-metallicity 40 M$_{\sun}$ star with the kinetic energy of explosion of 10$^{51}$ erg \citep{No06}.}
\label{fig1}
\end{figure}

\clearpage


\begin{figure}
\epsscale{1.0}
\plotone{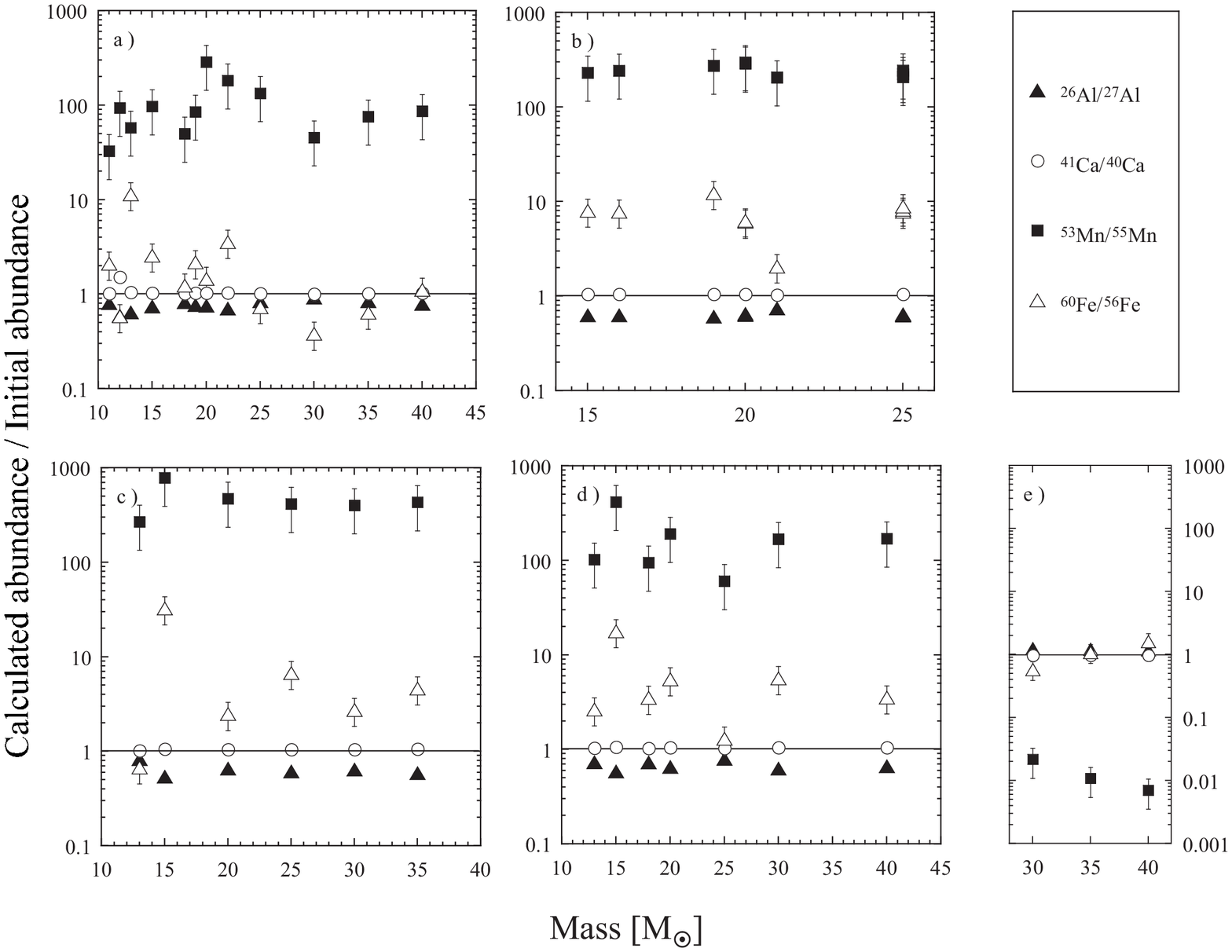}
\caption{Calculated initial abundances of $^{26}$Al, $^{41}$Ca, $^{53}$Mn, and $^{60}$Fe for non-fallback supernovae (\textit{a}-\textit{d}) and fallback supernovae (\textit{e}), normalized to their estimated initial abundances in the solar system, as a function of the stellar mass.  ($a$) Nucleosynthesis models for non-fallback supernovae by \citet{WW95}.  (\textit{b}) \cite{Ra02}.  (\textit{c}) \cite{CL04}.  (\textit{d}) \cite{No06}.  (\textit{e}) Nucleosynthesis models for fallback supernovae (based on results of \cite{WW95}).  Details of nucleosynthesis models such as the initial mass cut and the kinetic energy of explosion can be found in references listed above.  Error bars represent uncertainties of initial abundances of SLRs estimated from meteorites (see text).}
\label{fig2}
\end{figure}

\clearpage

\begin{figure}
\epsscale{1.0}
\plotone{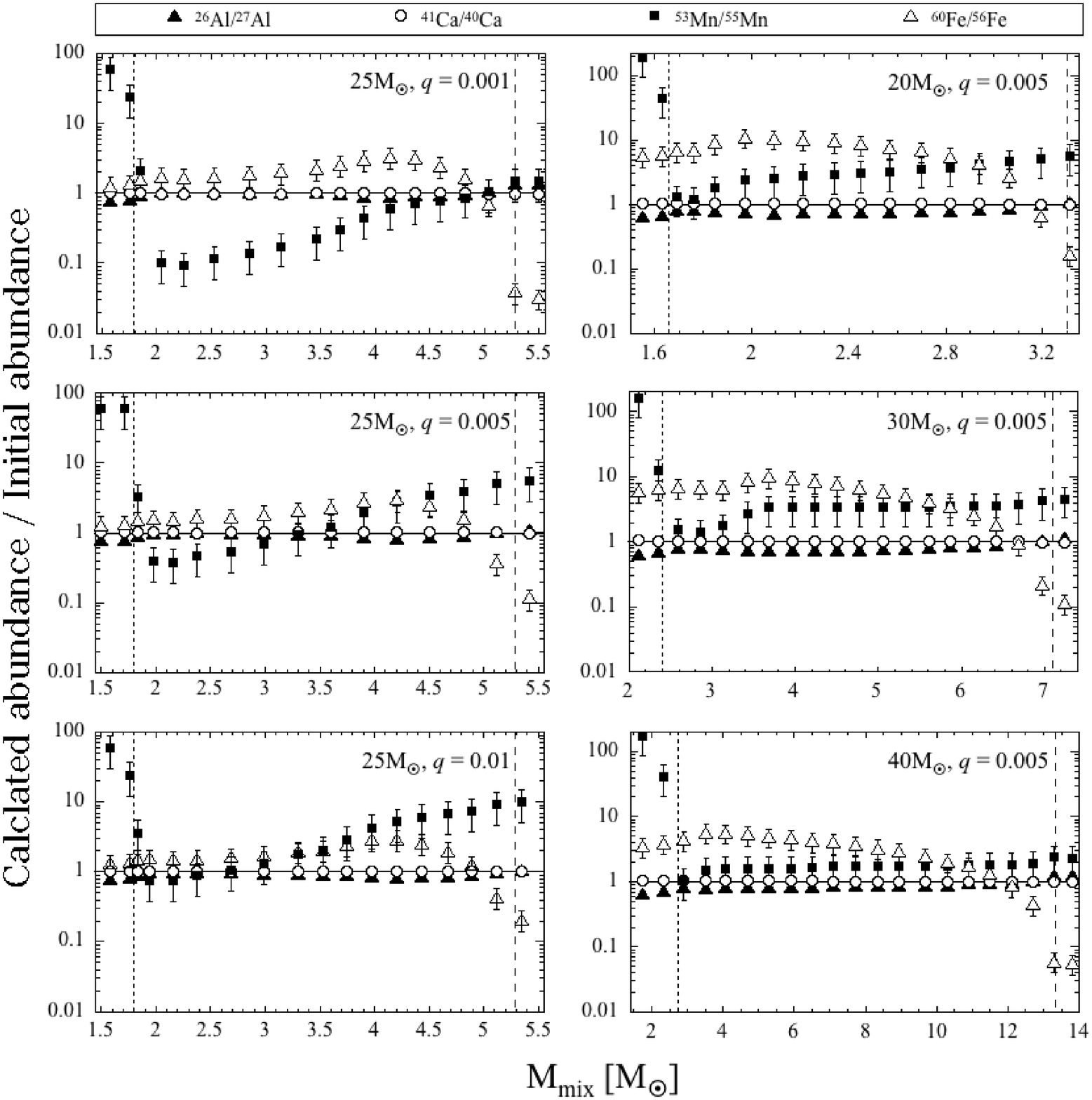}
\caption{Calculated initial abundances of $^{26}$Al, $^{41}$Ca, $^{53}$Mn, and $^{60}$Fe for supernovae with mixing-fallback having various masses as a function of \textit{M}$_{\rm mix}$ and \textit{q}.  The calculated abundances are normalized to their estimated initial abundances in the solar system.  The dilution factor $f_{0}$ ranges from 7$\times$10$^{-5}$ to 2$\times$10$^{-3}$, and time interval $\Delta$ ranges from 0.8 to 1.1 Myr. Each panel displays the results of 18-20 sets of calculations and the horizontal axis gives the mass of material in solar masses inside the outer boundary of the mixing region for each calculation. Dotted and dashed lines show the boundaries between the Si-burning and C/O-burning layers and between the C/O-burning and the He-burning layers respectively.}
\label{fig3}
\end{figure}

\clearpage

\begin{figure}
\epsscale{.70}
\plotone{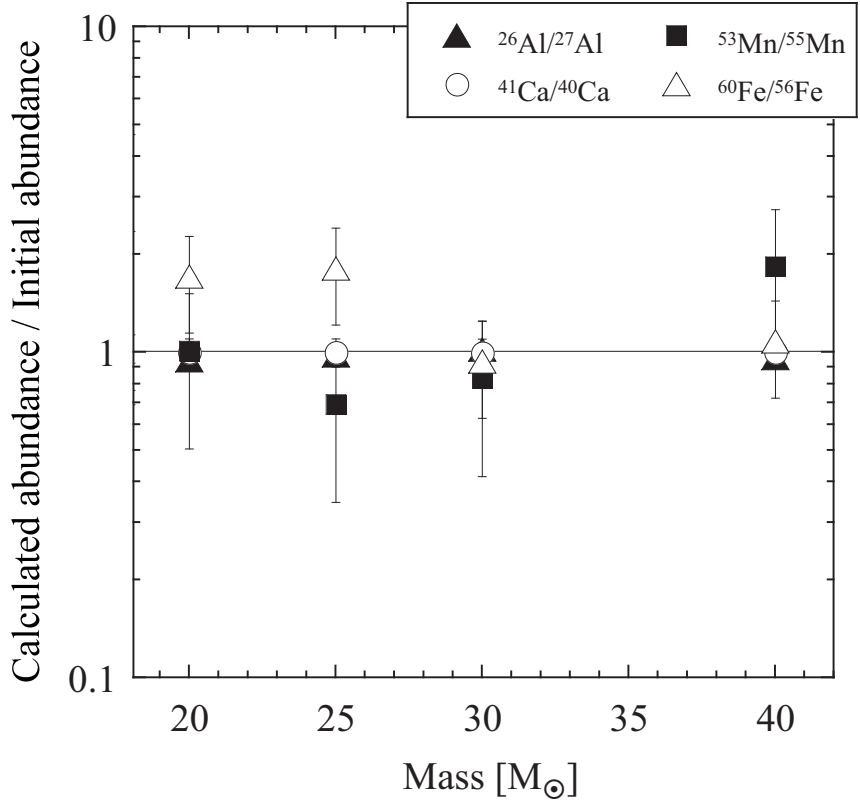}
\caption{The best estimates for the abundances of $^{26}$Al, $^{41}$Ca, $^{53}$Mn, and $^{60}$Fe in models for faint supernovae with mixing-fallback.  The values for the parameters are: \textit{q}=0.001, \textit{M}$_{\rm mix}$=3.1 M$_{\sun}$ (6.7$\times$10$^{8}$ K; peak temperature of the shock wave), \textit{f}$_{\rm 0}$=1.90$\times$10$^{-3}$, and $\Delta$=1.07 Myr for 20 M$_{\sun}$, \textit{q}=0.005, \textit{M}$_{\rm mix}$=3.0 M$_{\sun}$ (1.4$\times$10$^{9}$ K), \textit{f}$_{\rm 0}$=1.34$\times$10$^{-4}$, and $\Delta$=0.83 Myr for 25 M$_{\sun}$, \textit{q}=0.001, \textit{M}$_{\rm mix}$=6.7 M$_{\sun}$ (7.8$\times$10$^{8}$ K), \textit{f}$_{\rm 0}$=4.35$\times$10$^{-4}$, and $\Delta$=0.87 Myr for 30 M$_{\sun}$, and \textit{q}=0.005, \textit{M}$_{\rm mix}$=11.8 M$_{\sun}$ (5.8$\times$10$^{8}$ K), \textit{f}$_{\rm 0}$=1.64$\times$10$^{-4}$, and $\Delta$=0.75 Myr for 40 M$_{\sun}$.  Error bars represent uncertainties of initial abundances of SLRs estimated from meteorites (see text).}
\label{fig4}
\end{figure}

\clearpage


\begin{thebibliography}{}
\bibitem[Adams \& Laughlin(2001)]{AL01}Adams, F. C., \& Laughlin, G. 2001, \icarus, 150, 151
\bibitem[Anders \& Grevesse(1989)]{AG89}Anders, E., \& Grevesse, N. 1989, Geochim. Cosmochim. Acta, 53, 197
\bibitem[Arnould et al.(2006)]{Ar06}Arnould, M., Goriely, S., \& Meynet, G. 2006, \aap, 453, 653
\bibitem[Bizzarro et al.(2007)]{Bi07}Bizzarro, M., Ulfbeck, D., Trinquier, A., Thrane, K., Connelly, J. N., \& Meyer, B. S. 2007, Science, 316, 1178
\bibitem[Busso et al.(2003)]{Bu03}Busso, M., Gallino, R., \& Wasserburg, G. J. 2003, Publ. Astron. Soc. Australia, 20, 356
\bibitem[Chieffi \& Limongi(2004)]{CL04}Chieffi, A., \& Limongi, M. 2004, \apj, 608, 405
\bibitem[Goswami and Vanhala (2000)]{GV00}Goswami, J. N., \& Vanhala, H. A. T. 2000, in Protoatars and Planets IV, ed. V. Mannings, A. P. Boss, \& S. S. Russell (Tucson: Univ. Arizona Press), 963
\bibitem[Hsu et al.(2006)]{Hsu06}Hsu, W., Guan, Y., Leshin, L. A., Ushikubo, T., \& Wasserburg, G. J. \apj, 640, 525
\bibitem[Iwamoto et al.(2005)]{Iw05}Iwamoto, N., Umeda, H., Tominaga, N., Nomoto, K., \& Maeda, K. 2005, Science, 309, 451
\bibitem[Jacobsen(2005)]{Ja05}Jacobsen, S. B. 2005, in ASP Conf. Ser., 341, Chondrites and the Protoplanetary Disk, ed. A. N. Krot, E. R. D. Scott, \& B. Reipurth (San Francisco: ASP), 548
\bibitem[Kastner \& Meyers(1994)]{KM94} Kastner, J. H., \& Meyers, P. C. 1994, \apj, 421, 605
\bibitem[Kita et al.(2005)]{Ki05} Kita, N. T., Huss, G. R., Tachibana, S., Amelin, Y., Nyquist, L. E., \& Hutcheon, I. D. 2005, in ASP Conf. Ser. 341, Chondrites and the Protoplanetary Disk, ed. A. N. Krot, E. R. D. Scott, \& B. Reipurth (San Francisco: ASP), 558
\bibitem[Kroupa(2001)]{Kr01}Kroupa, P. 2001, MNRAS, 322, 231
\bibitem[Limongi \& Chieffi(2006)]{LC06}Limongi, M., \& Chieffi, A. 2006, \apj, 647, 483
\bibitem[McKeegan et al.(2000)]{Mc00}McKeegan, K. D., Chaussidon, M., \& Robert, F. 2000, Science, 289, 1334
\bibitem[McKeegan \& Davis(2003)]{McD03}McKeegan, K. D., \& Davis, A. M. 2003, in Meteorites, Comets, and Planets, ed. Davis, A. M., Vol. 1 Treatise on Geochemistry, eds. Holland, H. D., \& Turekian, K. K.  (Oxford: Elsevier), 431
\bibitem[Meyer(2005)]{Me05}Meyer, B. S. 2005, in ASP Conf. Ser. 341, Chondrites and the Protoplanetary Disk, ed. A. N. Krot, E. R. D. Scott, \& B. Reipurth (San Francisco: ASP), 515
\bibitem[Meyer \& Clayton(2000)]{MC00}Meyer, B. S., \& Clayton, D. D. 2000, Space Sci. Rev., 92, 133
\bibitem[Mostefaoui et al.(2005)]{Mo05}Mostefaoui, S., Lugmair, G. W., \& Hoppe, P. 2005 \apj, 625, 271
\bibitem[Nomoto et al.(2006)]{No06}Nomoto, K., Tominaga, N., Umeda, H., Kobayashi, C., \& Maeda, K. 2006, Nuclear Physics A (Specical Issue on Nuclear Astrophysics), 777, 424
\bibitem[Ouellette et al.(2005)]{Ou05}Ouellette, N., Desch, S. J., Hester, J. J., \& Leshin, L. A. 2005, in ASP Conf. Ser., 341, Chondrites and the Protoplanetary Disk, ed. A. N. Krot, E. R. D. Scott, \& B. Reipurth (San Francisco: ASP), 527
\bibitem[Rauscher et al.(2002)]{Ra02}Rauscher, T., Heger, A., Hoffman, R. D., \& Woosley, S. E. 2002, \apj, 576, 323
\bibitem[Sahijpal \& Soni(2006)]{SS06}Sahijipal, S., \& Soni, P. 2006, Meteorite \& Planet. Sci., 41, 953
\bibitem[Tachibana et al.(2006)]{Ta06}Tachibana, S., Huss, G. R., Kita, N. T., Shimoda, G. \& Morishita, Y. 2006, \apj, 639, L87
\bibitem[Tachibana \& Huss(2003)]{TH03}Tachibana, S., \& Huss, G. R. 2003, \apj, 588, L41
\bibitem[Tominaga et al.(2007)]{To07} Tominaga, N., Umeda, H., \& Nomoto, K. 2007, \apj, 660, 516
\bibitem[Umeda \& Nomoto(2002)]{UN02} Umeda, H., \& Nomoto, K. 2002, \apj, 565, 385 
\bibitem[Umeda \& Nomoto(2003)]{UN03} Umeda, H., \& Nomoto, K. 2003, \nat, 422, 871
\bibitem[Umeda \& Nomoto(2005)]{UN05} Umeda, H., \& Nomoto, K. 2005, \apj, 619, 427
\bibitem[Vanhara \& Boss(2002)]{VB02}Vanhara, H. A. T., \& Boss, A. P. 2002, \apj, 575, 1144 
\bibitem[Wasserburg et al.(2006)]{Wa06}Wasserburg, G. J., Busso, M., Gallino, R., \& Nollett, K. M. 2006, Nuclear Physics A, 777, 5
\bibitem[Woosley \& Weaver(1995)]{WW95}Woosley, S. E., \& Weaver, T. A. 1995, \apj, 101, 181
\end{thebibliography}
\end{document}